\documentclass[
 aip,
 amsmath,amssymb,
preprint
]{./revtex4-2}

\usepackage{natbib}
\usepackage{graphicx}
\usepackage{dcolumn}
\usepackage{bm}
\usepackage{multirow}
\usepackage{amsmath}
\usepackage{makecell}
\usepackage{tikz}

\usepackage[utf8]{inputenc}
\usepackage[T1]{fontenc}
\usepackage{mathptmx}

\begin{document}

\preprint{PoF/manuscript}

\title{Active flow control over a finite wall-mounted square cylinder by using multiple plasma actuators}

\author{Mustafa Z. Yousif} 
\email[]{These authors contributed equally to this work.}
\affiliation{School of Mechanical Engineering, Pusan National University, 2, Busandaehak-ro 63beon-gil, Geumjeong-gu, Busan, 46241, Republic of Korea}

\author{Yifan Yang}
\email[]{These authors contributed equally to this work.}

\author{Haifeng Zhou}
\author{Linqi Yu}
\author{Meng Zhang}

\author{Hee-Chang Lim}
\email[]{Corresponding author, hclim@pusan.ac.kr}
\thanks{}
\affiliation{School of Mechanical Engineering, Pusan National University, 2, Busandaehak-ro 63beon-gil, Geumjeong-gu, Busan, 46241, Republic of Korea}

\date{\today}

\begin{abstract}
The present study aims to investigate the effectiveness of plasma actuators in controlling the flow around a finite wall-mounted square cylinder (FWMSC) with a longitudinal aspect ratio of 4. The test is conducted in a small-scale closed return-type wind tunnel. The Reynolds number (${Re}_d$) of the experiments is 500 based on the width of the bluff body and the freestream velocity. The plasma actuators are installed on the top surface and the rear surface of the square cylinder. The induced flow velocities of the plasma actuators are modulated by adjusting the operating voltage and frequency of the high-voltage generator. In this work, particle image velocimetry (PIV) is used to obtain the velocity fields. Furthermore, force measurements are conducted to investigate the effect of using plasma actuators with different driving voltages on the drag force. Our results show that the plasma actuators can successfully suppress flow separation and reduce the size of the recirculation region and turbulent kinetic energy (TKE) in the wake. A correlation between the drag coefficient and the operating voltage of the power generator is also revealed and the mean drag coefficient is found to decrease with increasing imposing voltage. The plasma actuators can enhance the momentum exchange and the interactive behavior between the shear layer and the flow separation region, resulting in flow reattachment at the free end and shrinkage of the recirculation zone in the near-wake region of the bluff body. Overall, the present study demonstrates the practical effectiveness of using plasma actuators for active flow control around FWMSC.

\end{abstract}

\maketitle

\section{Introduction}\label{sec:introduction}
The flow around a finite wall-mounted square cylinder (FWMSC) has been extensively studied over several decades due to its wide range of applications in numerous fields. This model provides a simplified representation of common obstacles found on vehicles and aircrafts, such as rear-view mirrors and landing gears. In recent years, significant progress has been made in understanding the complicated three-dimensional (3-D) flow structure around a FWMSC, with numerous studies conducted in this area. The flow structure is composed of four primary types of vortices, namely the tip vortex, spanwise vortex, horseshoe vortex and base vortex. The tip vortex is caused by the downwash flow from the free end of the FWMSC, while the base vortex is induced by the upwash flow from the boundary layer. The thickness of the boundary layer can strengthen the base vortex.\cite{dasilva2020,uffinger2013} By increasing the submerged length of the bluff body in the boundary layer, the mean drag force and fluctuation lift force will be reduced.\cite{wang2018} Furthermore, the tip and the spanwise vortices will merge near the free end and form an arch-type flow structure.\cite{wang2009}\par

The aspect ratio $H/d$ where $H$ and $d$ represent the height and width of the FWMSC, respectively, plays a significant role in spanwise vortex shedding structures. Specifically, the aspect ratio has a considerable influence on the type of spanwise vortex shedding. When the aspect ratio is smaller than the critical value, the downwash flow becomes relatively strong, resulting in a symmetric type of spanwise vortex. However, under these circumstances, the lift coefficient has high amplitude but is not periodic. In contrast, when the aspect ratio is above the critical value, the downwash flow becomes weak and cannot dominate the wake structure. Consequently, both symmetric and asymmetric shedding occurs intermittently.\cite{bourgeois2012,pattenden2005,sattari2012,wang2022,yousif2021} Although the amplitude of the lift coefficient decreases, it becomes more periodic compared to the previous case. In general, the critical aspect ratio is influenced by the oncoming flow conditions, including the Reynolds number, turbulence intensity and boundary layer thickness in the bluff body region\cite{behera2019,okamoto1992}, and a suggested value of 2.5 was given for the finite square cylinder.\cite{sakamoto1983} Our profound understanding of the flow structure near the finite-length bluff body enables us to employ methods to control the flow around it, such as suppression of flow separation and drag reduction. When it comes to a square cylinder, flow control can generally be divided into two types depending on the input energy: passive and active control. In addition, active flow control can be further classified into closed-loop control and open-loop control due to the different feedback mechanisms. \cite{cattafesta2011}\par

Compared to active flow control, studies on passive flow control are more commonly conducted due to its easy of implementation. In particular, passive flow control methods require no additional energy input, rendering them more cost-effective than their active counterparts. Consequently, a vast number of studies on passive flow control methods have been conducted during the past few decades. These methods primarily involve surface modifications, such as roughness, splitter plates and holes, as well as the use of external elements and geometric modifications in the spanwise direction. Sharma \textit{et al.}\cite{sharma2020} have experimentally studied flow control by using a rigid and flexible splitter installed in the rear face of the finite bluff body at a low Reynolds number. The authors found that the wake frequency and mean drag coefficient varied non-monotonically with the splitter plate length, which is an important parameter because it can achieve flow control and significantly alter the wake. Furthermore, Rinoshika and Rinoshika\cite{rinoshika2019} experimentally studied passive flow control by drilling an inclined hole in a wall-mounted short square cylinder from the front surface side to its free end. The inclined hole decreased the separation bubble area near the free end and the rear regions and could more efficiently suppress the vortex and Reynolds shear stress. Wang \textit{et al.}\cite{wang2022} conducted an experimental study to investigate the control effect of a flexible plate attached at the free end. The experiments were carried out with Reynolds number ranging from 10,960 to 54,800. The results indicated that the control effectiveness was directly influenced by the length of the flexible plate, with a substantial reduction in the aerodynamic force occurring when the freestream velocity exceeded the critical value. Specifically, a maximum reduction in the mean drag of approximately 5$\%$ was achieved.\par

Compared to passive flow control, active flow control offers numerous advantages. Passive flow control typically involves the modification of the target’s surface structure, which can sometimes be impractical or impossible to perform. Unlike passive flow control, active flow control seeks to introduce extra momentum into the flow field without modifying the surface structure of the designated model, making it a more practical and efficient approach. As a result, active flow control has gained significant attention in recent years as a promising technique for controlling aerodynamic forces and flow structures around finite wall-mounted square cylinders. Wang \textit{et al.}\cite{wang2018} installed a steady slot suction at its free end to control the aerodynamic force. The authors used different suction velocities to reduce the mean drag coefficient, fluctuating drag and lift coefficient, and reached a reduction of 3.6$\%$, 17.8$\%$ and 45.5$\%$, respectively. Li \textit{et al.}\cite{li2021} performed an experimental study on active flow control by using a dual synthetic jet on the free end of a finite FWMSC. To generate this synthetic jet, a piezoelectric dual synthetic jet actuator was installed at the leading edge of the free end. The experimental results demonstrated a clear correlation between the control efficiency and the amplitude and frequency of the dual synthetic jet. Moreover, the aerodynamic coefficient and free end shear flow were effectively suppressed, while the turbulent kinematic energy was significantly enhanced.\par

The Dielectric Barrier Discharge (DBD) plasma actuator has recently garnered significant attention due to its characteristics of flexibility, high efficiency and fast response.\cite{shen2020} A typical plasma actuator consists of two electrodes and a dielectric material between them. Application of high-voltage AC to the electrodes will ionise the air near them, which is subsequently accelerated by the electric field generated by the AC, generating a plasma wind. Numerous studies have been conducted on 2-D bluff bodies. For example, Chen and Wen \cite{chen2021} have applied the three different types of plasma actuators on the D-shaped bluff body, while
Zhu \textit{et al.}\cite{Zhu2021} numerically studied the performance of installing plasma actuators at different locations on an infinite square cylinder. All previous studies showed that plasma actuators could perform well in suppressing vortex shedding and reducing the aerodynamic coefficient.\par

The present study aims to investigate the feasibility of using plasma actuators for active flow control on the FWMSC. Multiple plasma actuators were installed on different surfaces of the FWMSC to enhance our understanding on the effect of plasma actuators on 3-D bluff body flows. Three different configurations of plasma actuators are compared to better comprehend the mechanism of plasma actuator control for various shedding types.\par
The remainder of this paper is organised as follows: Section.~\ref{sec:Experimental Setup} introduces the experimental design of the model, wind tunnel, plasma actuators and measurement equipments used in the present study. The results, including contours, profiles, statistics and aerodynamic coefficients, are presented in section.~\ref{sec:Results}. Finally, section.~\ref{sec:Conclusions} provides conclusions based on the results of our study.\par

\section{Experimental Setup}\label{sec:Experimental Setup}

The experiments are conducted in a return-type wind tunnel with a test section of 250 mm $\times$ 250 mm $\times$ 1000 mm. The experimental facility comprises a FWMSC made of an acrylic board, plasma actuators, a signal generator, a voltage amplifier, a load cell for force measurement, a pitot tube for point velocity measurement and a particle image velocity (PIV) system for velocity measurements.\par

\begin{figure}
\centering 
\includegraphics[angle=0, trim=0 0 0 0, width=1\textwidth]{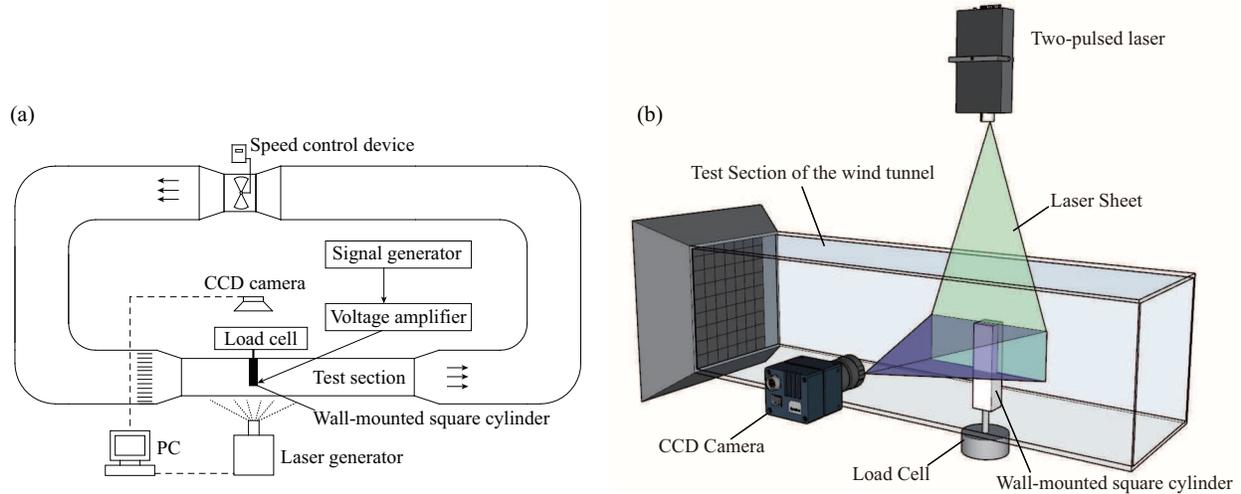}
\caption[]{(a) Schematic diagram of wind tunnel and experimental equipments. (b) Schematic diagram of the wind tunnel test section.}
\label{fig:1-schematic of wind tunnel}
\end{figure}

A pitot tube is placed upstream of the model to measure the freestream velocity. The freestream velocity $U_\infty$, is approximately 0.4 $m/s$, so the corresponding ${Re}_d$ based on the width $d$ of the square cylinder is approximately 500. The corresponding turbulence intensity of the freestream is less than 0.8$\%$. The schematic diagram of wind tunnel and the arrangement of experimental equipments are shown in Fig.~\ref{fig:1-schematic of wind tunnel}(a), the schematic diagram of the wind tunnel test section is shown in Fig.~\ref{fig:1-schematic of wind tunnel}(b).\par

\subsection{Model geometry and the arrangement of plasma actuators}

The experimental model used in this study is a FWMSC mounted on the wall of the wind tunnel's test section. The model, composed of acrylic material, had a width ($d$) of 20 mm and a height ($h$) of 80 mm. The blockage ratio of the model is approximately 2.6$\%$. The model consisted of two plasma actuators installed on its respective surfaces, with plasma actuator 1 located on the top surface and plasma actuator 2 on the rear surface. Detailed configurations of the model are provided in Fig.~\ref{fig:2-arrangement of PA}. Both the exposed and encapsulated electrodes of the plasma actuators are made of copper, with a thickness of 0.07mm. The dielectric material is composed of two layers of Kapton tape, each with a thickness of 0.065 mm, resulting in a total thickness of 0.13 mm. Since this thickness is significantly small compared to the model's width, the impact of the plasma actuators on the flow field could be considered negligible. The plasma actuators are operated by imposing high-voltage AC on the exposed electrodes. The power has a sinusoidal waveform with a frequency of 900 Hz and a peak-to-peak voltage of 9 kV. The AC power is generated by a function generator (Agilent 33522A) and then amplified by a voltage amplifier (Trek MODEL 610E) for plasma actuator operation. Because the natural vortex shedding frequency is much smaller than the input AC signal frequency, the flow control is considered continuous.\cite{shen2016}\par

In this work, except for the no-control case, three different cases are experimentally considered to investigate the plasma actuator flow control. In case 1, only actuator 1 is activated, while actuator 2 is turned off. In case 2, actuator 1 is turned off and actuator 2 is turned on. Finally, in case 3, both actuators are turned on.\par

\subsection{Velocity measurements}

\begin{figure}
\centering 
\includegraphics[angle=0, trim=0 0 0 0, width=1\textwidth]{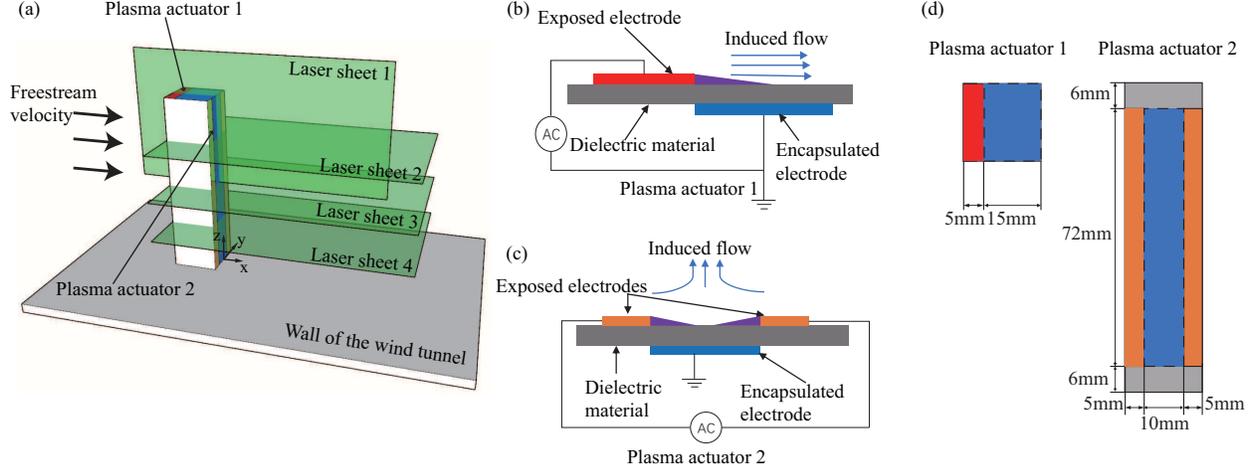}
\caption[]{(a) Arrangement of plasma actuators and laser sheets. (b, c) Configuration of the two plasma actuators. (d) Top view (left) of the plasma actuator 1 and rear view of plasma actuator 2 (right).}
\label{fig:2-arrangement of PA}
\end{figure}

Our experimental process use the PIV to obtain the flow field around the bluff body. The PIV system comprises a CCD camera (VC-12MX) with 4096 $\times$ 3072 pixels resolution and a two-pulsed laser (Evergreen, EVG00070). In the wind tunnel experiments, the wind tunnel is seeded by olive oil droplets generated by the TSI 9307 particle generator. The average size of the oil droplet is 0.65 $\mu$m. The planes chosen in this experiment are shown in Fig.~\ref{fig:2-arrangement of PA}(a). Along the spanwise direction, three planes are arranged within $z/d$ = 1, 2 and 3 at the rear region of the bluff body, and we also considered the plane location where $y/d$ = 0. For each location, 2000 snapshots are recorded for post-processing.\par

\subsection{Drag force measurement}

A load cell (DACELL-MC15) is used to measure the drag force in this experiment. Fig.~\ref{fig:1-schematic of wind tunnel}(b) demonstrates the schematics of the load cell arrangement in the test section of the wind tunnel. To eliminate low-frequency vibrations caused by the wind tunnel, the load cell is mounted on a flat fixed frame placed on the ground. In addition, the load cell is connected to the bluff body using a metal cylinder.\par

\section{Results and discussion}\label{sec:Results}

\subsection{Visualisation of the near-wake flow structure}

\begin{figure}
\centering 
\includegraphics[angle=0, trim=0 0 0 0, width=1\textwidth]{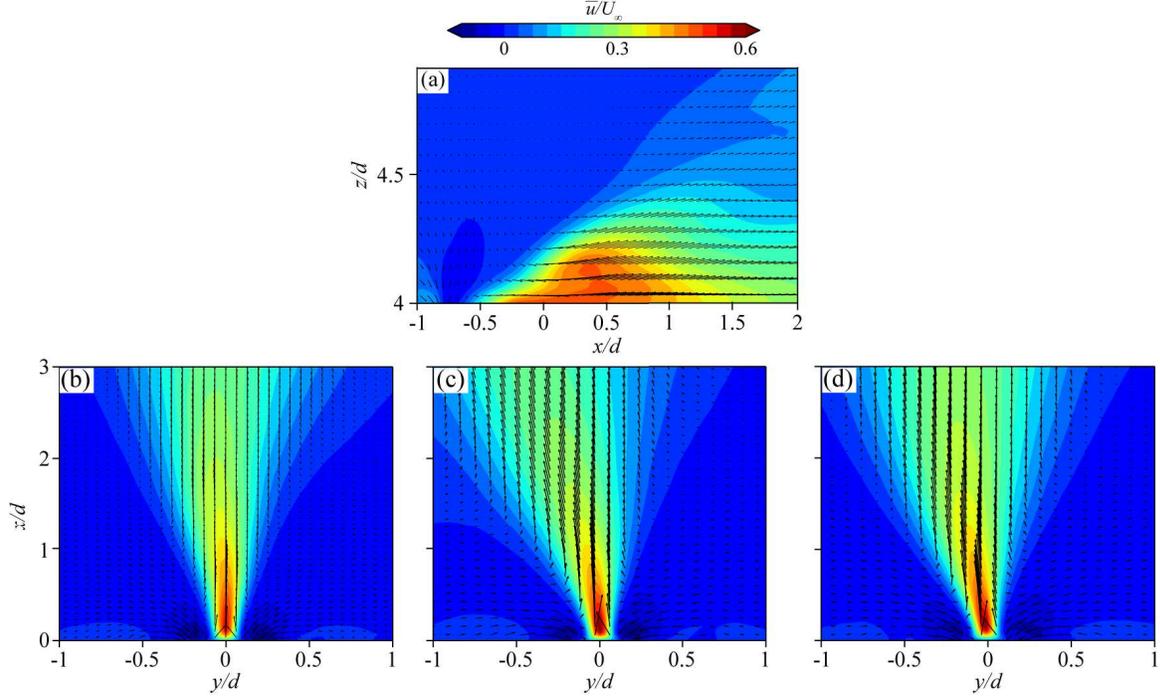}
\caption[]{The induced flow field in the quiescent air. (a) Induced flow field of plasma actuator 1 at $y/d$ = 0. (b, c, d) Induced flow field of plasma actuator 2 at $z/d$ = 1, 2 and 3, respectively.}
\label{fig:3-induced flow field}
\end{figure}

In this study, we investigate the induced flow field of plasma actuators to gain a deeper understanding of the flow control mechanisms. Figure.~\ref{fig:3-induced flow field}(a) shows that the induced flow from plasma actuator 1 is generated behind the exposed electrode edge and directed towards the encapsulated electrode. To further analyse the induced flow field, two-dimensional flow visualization is conducted in three different sections ($z/d$ = 1, 2 and 3), as shown in Fig.~\ref{fig:3-induced flow field}(b, c and d). Two single plasma actuators (actuator 2) are positioned at the back of the model, and the induced flow from each actuator is generated at the edge of the respective actuator and directed in opposite directions. The induced flows collide in the centreline behind the model, creating a mixed plasma jet towards the downstream direction ($x$ direction). Here, it is important to note that the induced flow velocities are nearly constant along the wall-normal direction. However, the jet flow is slightly deflected due to a main flow disturbance and improper configuration of the two exposed electrodes, such as a non-parallel alignment.\par

\begin{figure}
\centering 
\includegraphics[angle=0, trim=0 0 0 0, width=1\textwidth]{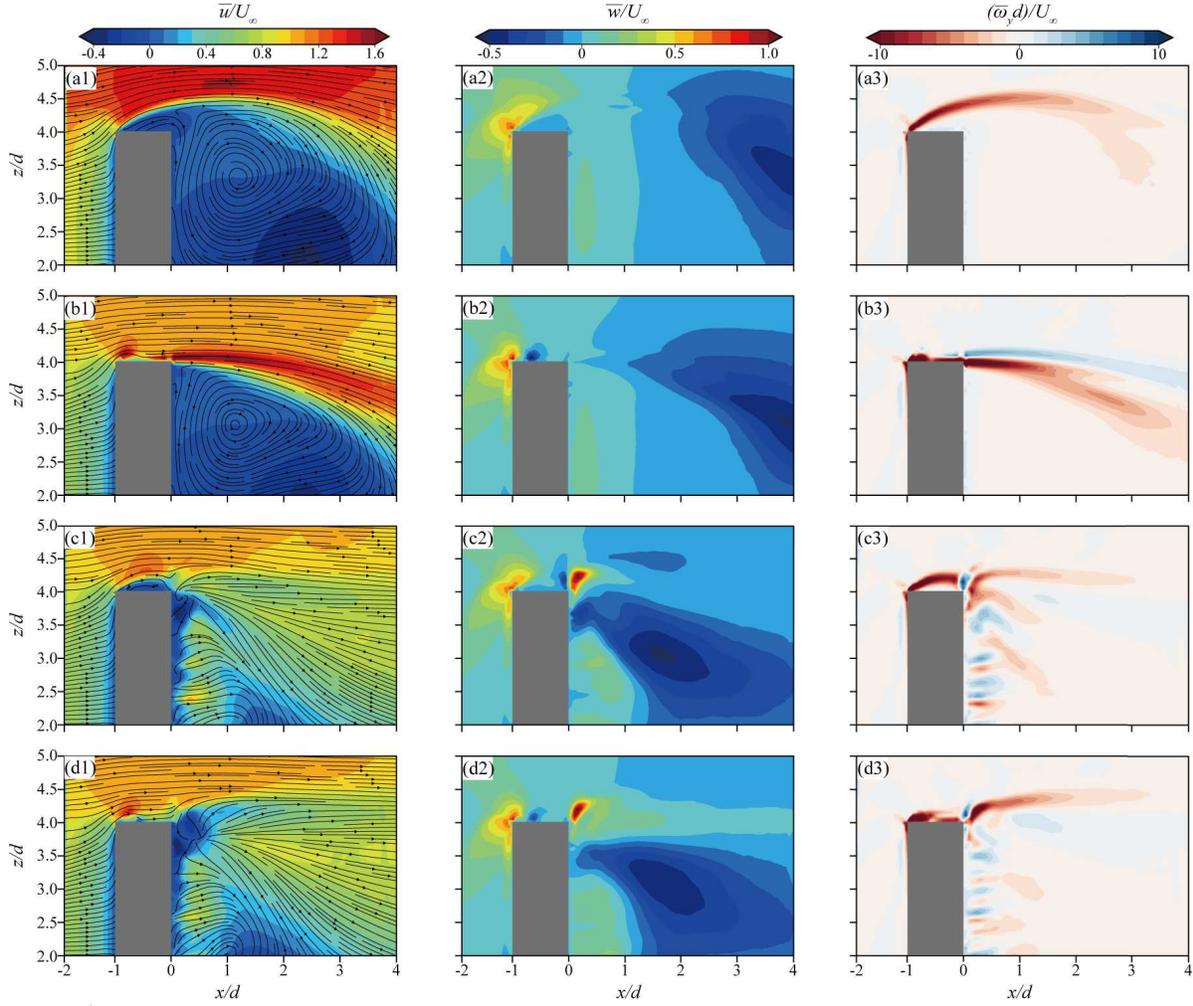}
\caption[]{Time-averaged contours of the dimensionless streamwise velocity and streamlines (left column), spanwise velocity (middle column) and vorticity (right column) for $y/d$ = 0. (a) No control; (b) case 1; (c) case 2; and (d) case 3.}
\label{fig:4-sideview}
\end{figure}

The control efficiency of the plasma actuators could be reflected by the suppression of the free end and the development of spanwise vortex shedding patterns. Figure~\ref{fig:4-sideview} exhibits the time-averaged contours of the dimensionless streamwise velocity, wall-normal velocity and vorticity in the central lateral plane ($y/d$ = 0). As shown in Fig.~\ref{fig:4-sideview}(a1–a3), the flow separation from the leading edge of the bluff body and a recirculation zone generated behind the bluff body. This phenomenon is similar to the flow structure reported by Saha\cite{saha2013} with respect to a wall-mounted square cylinder with $h/d$ = 4. \par

The results of the present study reveal interesting findings for case 1, where the reattachment flow occurs near the top surface of the bluff body, as illustrated in Fig.~\ref{fig:4-sideview}(b1–b3). The separated flow rapidly reattaches to the top surface of the bluff body, causing a change in the shape of the recirculation zone. Notably, the core of the recirculation bubble moves downwards from $z/d$ = 3.4 to $z/d$ = 3, indicating a significant impact on the free shear layer structures and downwash flow characteristics. \par 

For case 2 shown in Fig.~\ref{fig:4-sideview}(c1–c3), where actuator 1 is turned off and actuator 2 is turned on, our study identifies that the induced streamwise velocity can suppress the upwash flow near the wake region of the bluff body. In addition, the induced streamwise velocity could reduce the adverse flow region in the recirculation zone by injecting high velocity induced flow downstream, leading to an increased momentum exchange in the recirculation zone and a decrease in the adverse pressure gradient. Consequently, the recirculation zone is diminished, a behaviour that is also reflected by the deformation of the shear layer structure. \par

As shown in Fig.~\ref{fig:4-sideview}(c1–c3), a new shear layer is formed and a smaller recirculation region is formed behind the model. Furthermore, in the region where $z/d$ ranges between 3.5 and 4 there appears to be a strong upwash flow behind the trailing edge over the top surface of the body (i.e., $x/d$ = 0, $z/d$ = 4) in cases 2 and 3, as shown in Fig.~\ref{fig:4-sideview}(c2, d2). This seems to be attributed to the wall-normal pressure gradient of the new shear layer and the high-velocity region above the bluff body. A similar tendency is also observed in case 3, as shown in Fig.~\ref{fig:4-sideview}(d1–d3).\par

\begin{figure}
\centering 
\includegraphics[angle=0, trim=0 0 0 0, width=1\textwidth]{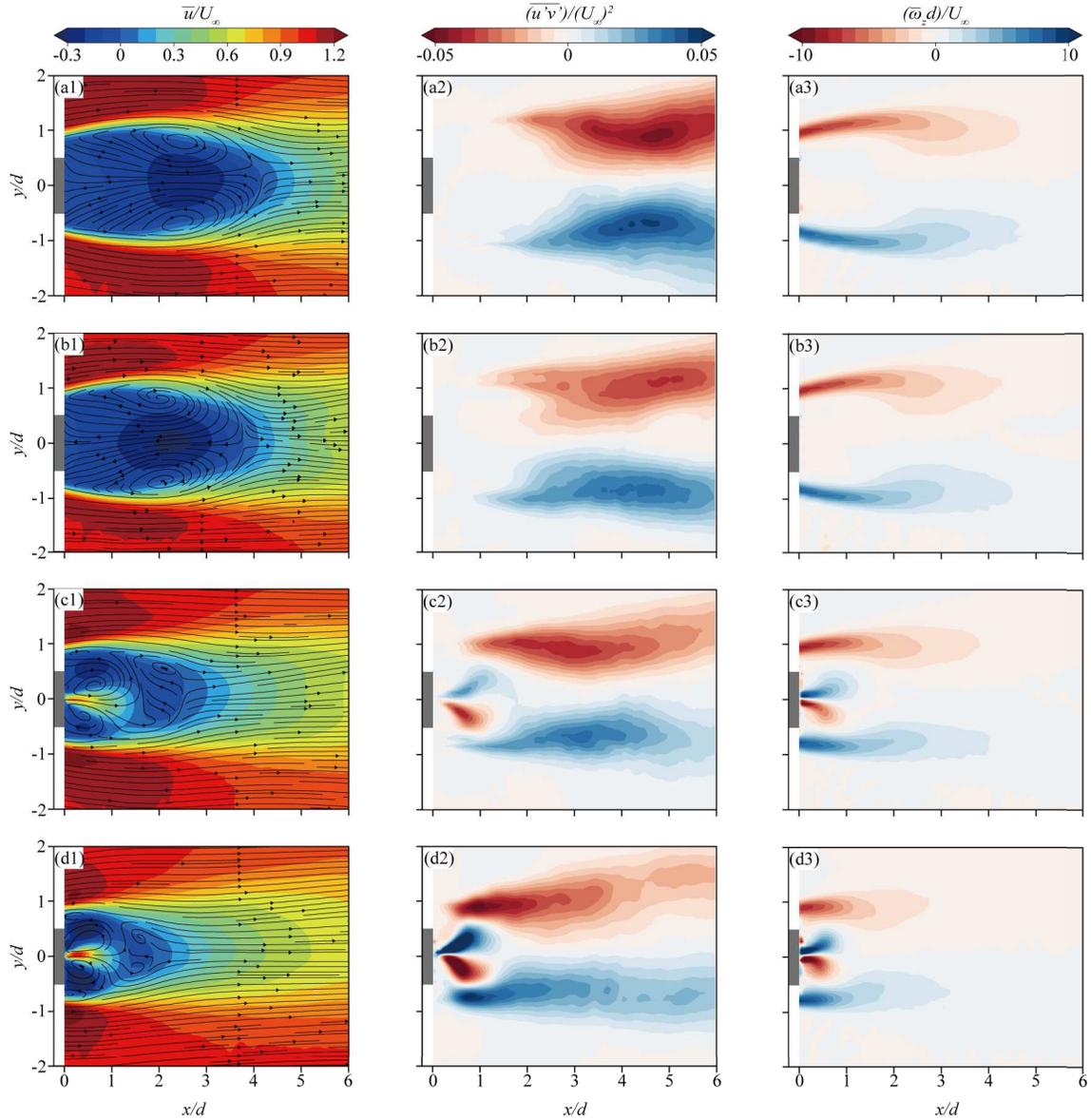}
\caption[]{Time-averaged contours of streamwise velocity and streamlines (left column), Reynolds stress component $\overline{u^\prime v^\prime}$ (middle column), and wall-normal vorticity (right column) for $z/d$ = 2. (a) No control; (b) case 1; (c) case 2; and (d) case 3.}
\label{fig:5-topview2d}
\end{figure}

Figure~\ref{fig:5-topview2d} presents the overall flow visualisation behind the model along the horizontal plane (i.e., $z/d$ = 2), indicating the time-averaged contours of streamwise velocity, streamlines, Reynolds shear stress and vorticity contours. For the no-control case (Fig. 5(a1–a3)), it can be conjectured that the adverse pressure gradient observed behind the bluff body leads to the formation of a wide recirculation zone, which could cause a substantial drag force acting on the bluff body (see Fig.~\ref{fig:5-topview2d}(a1)). As shown in Fig. 5(a2, a3), the strong variation of Reynolds stress affects wider wake region in the downstream of the bluff body, and the mean vorticity contour seems consistent along the centerline of the bluff body. This behavior would determine the range of vortex shedding frequency.\cite{chen2021}\par

In case 1, the wake flow field varies a little (see Fig.~\ref{fig:5-topview2d}(b1–b3)), and the maximum Reynolds stress is notably decreased compared with that obtained from the no-control case (see Fig.~\ref{fig:5-topview2d}(b2)). As shown in Fig.~\ref{fig:5-topview2d}(c1–c3), when the plasma actuator 2 is turned on, the wake flow structure reveals a significant variation. The recirculation zone in case 2 is suppressed by the control of plasma actuator 2, and the reattachment point moves upstream from $x/d$ = 4 to $x/d$ = 2.5. Furthermore, a substantial variation in the Reynolds stress and mean vorticity is also identified. As shown in Fig.~\ref{fig:5-topview2d}(c2), the maximum Reynolds stress moves upstream from $x/d$ = 4.5 to $x/d$ = 3. Our findings reveal that the use of a plasma actuator can effectively suppress velocity fluctuations in the wake, resulting in a substantial change in the wake structure, as shown in Fig.~\ref{fig:5-topview2d}(c3). The purpose is to inject high-momentum plasma into the flow field, leading to momentum exchange that accelerates the relatively slow velocity flow in the recirculation zone. As a result, the pressure gradient decreases, effectively suppressing flow separation. The shear layer tends to become narrower and deviate along the centreline behind the bluff body.\par

In case 3, the results (as shown in Fig.~\ref{fig:5-topview2d}(d1–d3)) indicate that the recirculation zone becomes smaller, leading to a further decrease in the Reynolds stress downstream. Additionally, the shear layer becomes narrower to both the centreline and the rear surface of the bluff body, demonstrating that case 3 is more efficient than case 2. It can be inferred that among the three cases, case 3 is the optimal control case because it can alter the wake structure significantly. In summary, our findings indicate that the plasma actuator on the rear surface plays a crucial role in suppressing the flow separation, inducing significant changes in the wake structure. In addition, the plasma actuator 1 on the top surface also contributes to reducing the turbulent wake.\par

\begin{figure}
\centering 
\includegraphics[angle=0, trim=0 0 0 0, width=1\textwidth]{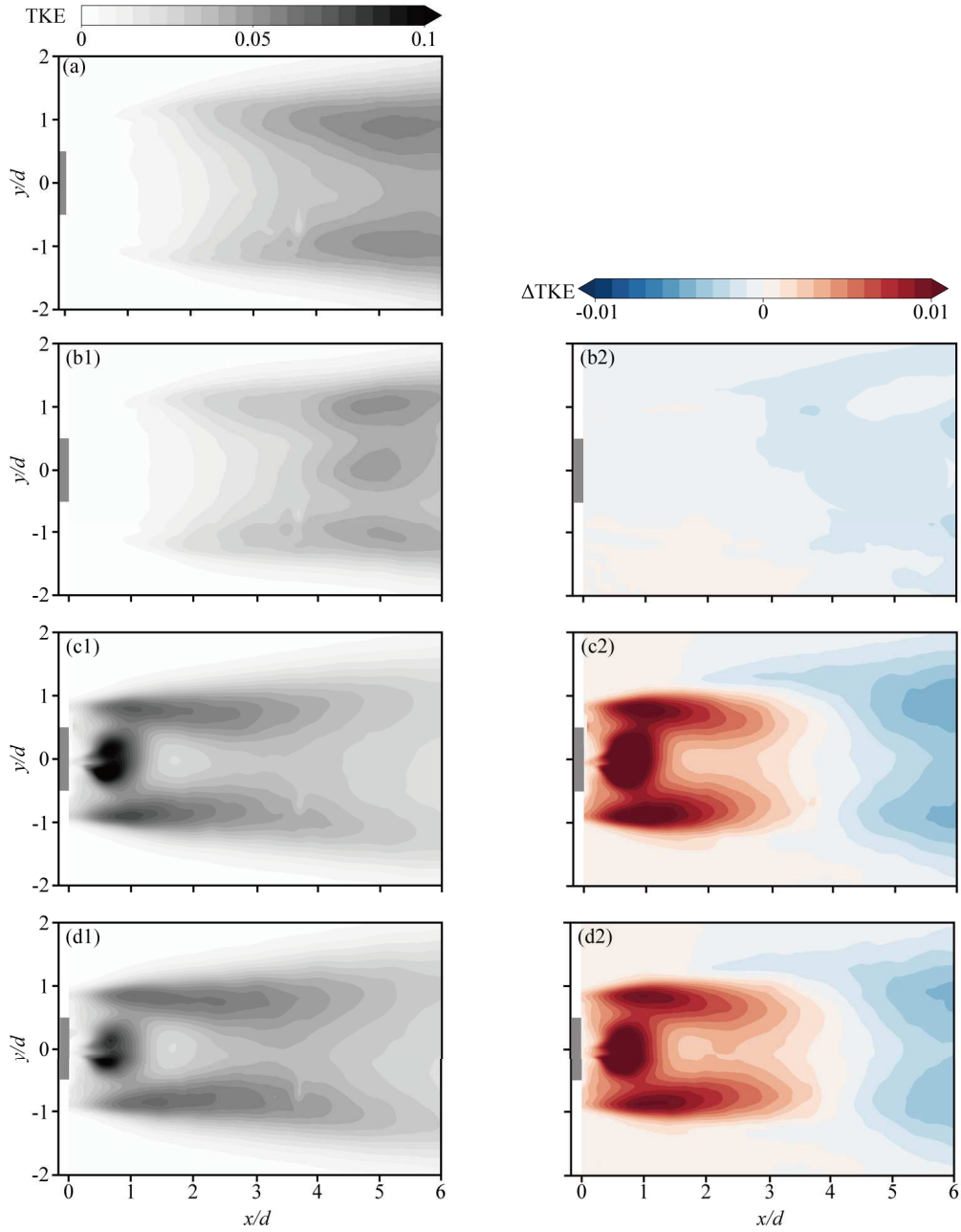}
\caption[]{TKE and $\Delta$TKE of the no-control and three control cases. (a) Uncontrolled case; (b1)(b2) case 1; (c1)(c2) case 2; and (d1)(d2) case 3.}
\label{fig:6-TKE}
\end{figure}

Figure 6 shows the contour of the turbulent kinetic energy (TKE) profile for each case investigated in the present study. Here, TKE is defined as:

\begin{equation} \label{eqn:eq1}
TKE = \frac{\overline{(u^\prime)^2}+\overline{(v^\prime)^2}}{2U^2_{\infty}},
\end{equation}

\noindent where $u'$ and $v'$ components are the streamwise and spanwise fluctuating velocity, respectively. Note here that TKE and $\Delta$TKE represent the level of fluctuating velocity and the difference in TKE between the control cases and the no-control case, respectively. In the no-control case, the value of TKE becomes big after reaching $x/d$ = 2 due to the strong vortex shedding behaviour. In contrast, the value of TKE becomes very low in the near-wake region between $x/d$ = 0 and $x/d$ = 2 due to the recirculation area. After activating the plasma actuator control, the maximum TKE tends to move upstream. As shown in Fig.~\ref{fig:6-TKE}(b1–d1, b2–d2), the plasma actuator increases the strength of TKE near the wake because of the induced turbulent flow, but the range of TKE from $x/d$ = 4 to $x/d$ = 6 becomes narrower compared with the no-control case. The results show that the actuator control reduces the level of turbulent energy in the downstream, indicating the suppression of the vortex shedding.\par

\begin{figure}
\centering 
\includegraphics[angle=0, trim=0 0 0 0, width=1\textwidth]{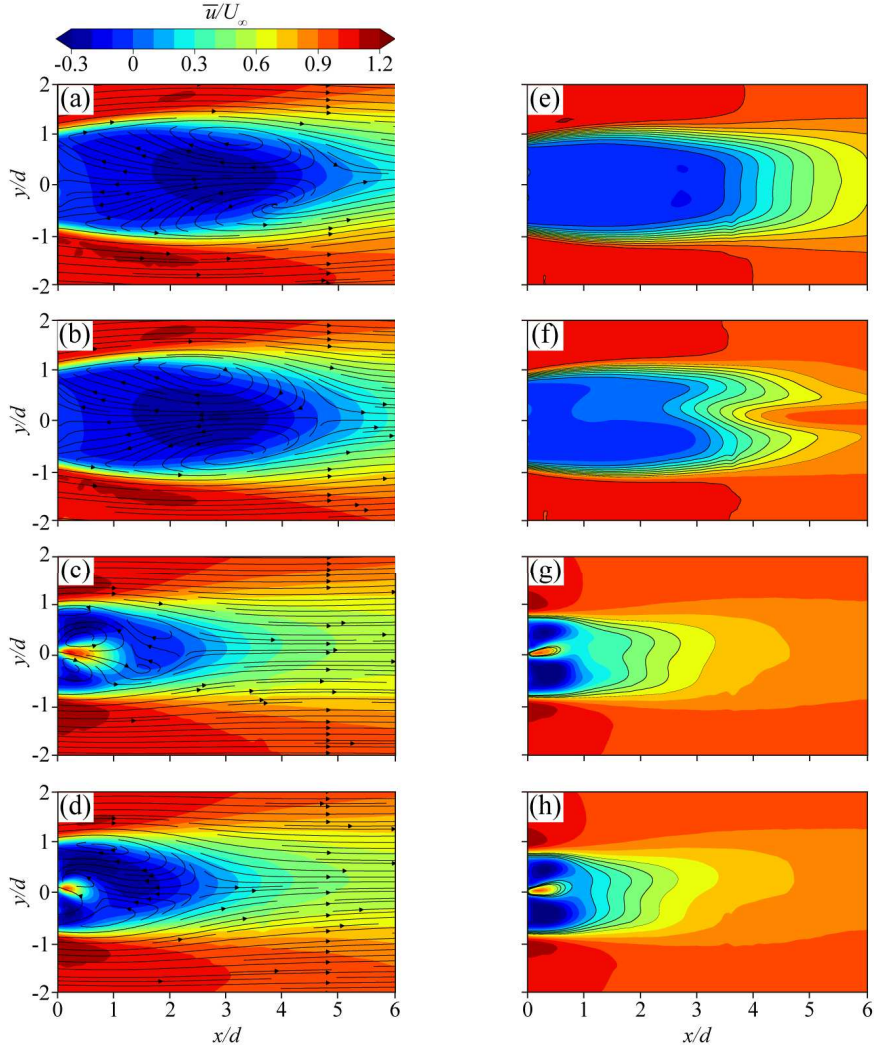}
\caption[]{Time-averaged contours of the streamwise velocity at $z/d$ = 1[(a)-(d)], $z/d$ = 3[(e)-(h)]. (a)(e) Uncontrolled case; (b)(f) case1; (c)(g) case 2; and (d)(h) case 3.}
\label{fig:7-topview1d3d}
\end{figure}

Figure~\ref{fig:7-topview1d3d} demonstrates the time-averaged contours of the dimensionless streamwise velocity at the planes where $z/d$ = 1 and $z/d$ = 3. The spanwise velocity fields at $z/d$ = 1 and $z/d$ = 3 are shown in Fig.~\ref{fig:7-topview1d3d}(a)–(d) and Fig.~\ref{fig:7-topview1d3d}(e)–(h), respectively, and it is evident that the overall flow structures are substantially different from one another. The recirculation zone at $z/d$ = 1 looks bigger than that at $z/d$ = 2 and the reattachment point appears around $x/d$ = 5, which is further behind the central plane than where the flow is reattached at $x/d$ = 4. This would be attributed to the separated upwash shear flow from the side of the model, which is also observed by Wang \textit{et al.}.\cite{wang2018} Figure~\ref{fig:7-topview1d3d}(e)–(h) exhibits the streamwise velocity field at $z/d$ = 3, which is close to the free end of the bluff body. The dark solid lines represent the isolines, which have an M-shape, indicating the faster recovery of the wind speed in the central region. Interestingly, as the flow control is made, the recirculation zone moves closer to the centreline of the bluff body, which could be attributed to the highly complicated three-dimensional wake flow near the free end. The free end vortex shedding would suppress the spanwise vortex shedding and the arch-type vortex would be formed by these two different types of vortex shedding near the free end. \cite{bourgeois2012,Wang2012,wang2009} Nevertheless, it is worth noting that the flow structure and the shape of vortex shedding are different at various heights. Our results indicate that the actuator control would be an effective way to suppress complicated vortex shedding structures.\par

\subsection{Statistical analysis of the wake flow}

\begin{figure}
\centering 
\includegraphics[angle=0, trim=0 0 0 0, width=1\textwidth]{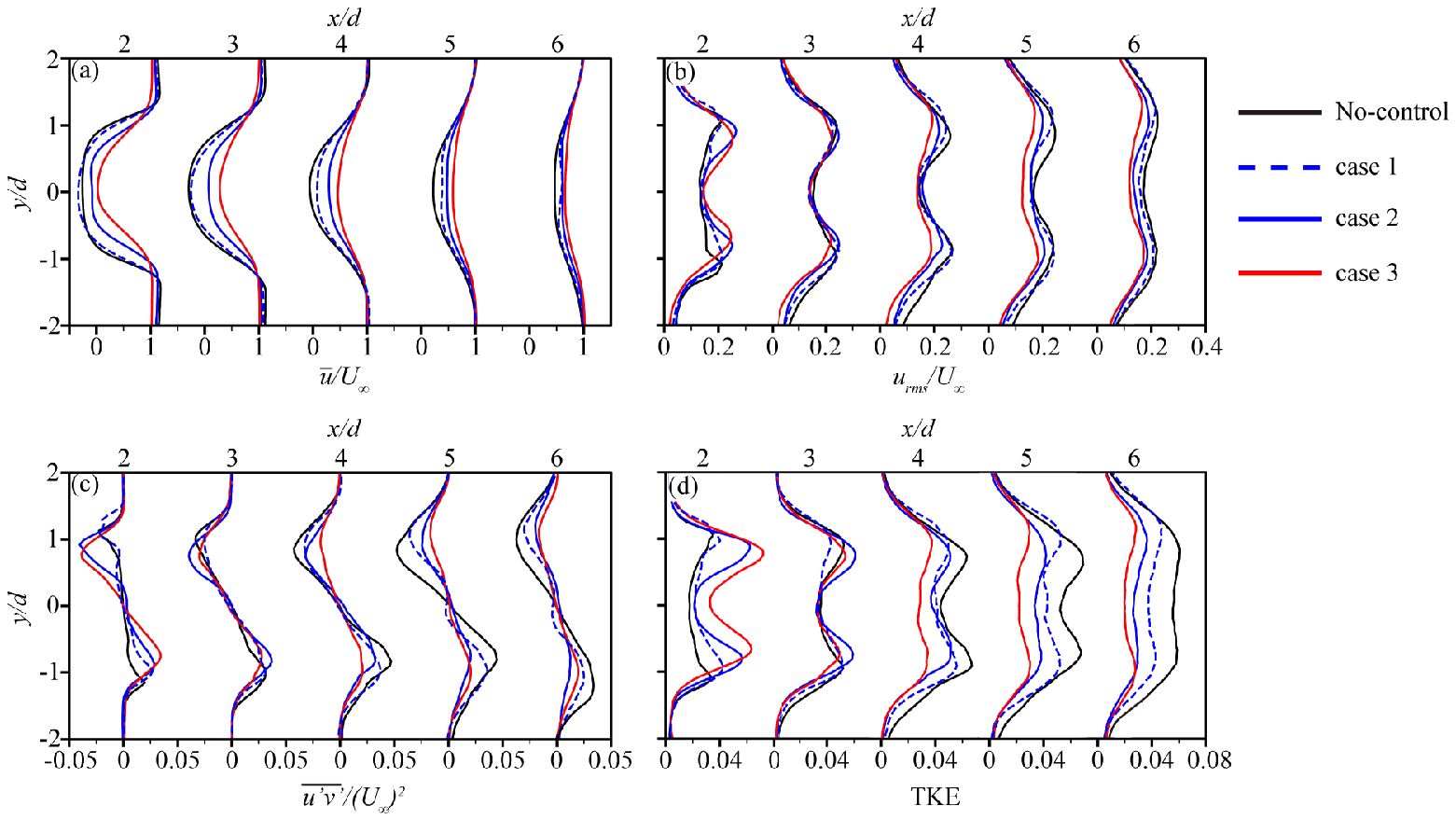}
\caption[]{Profiles of the dimensionless (a) time-averaged streamwise velocity; (b) root mean square of the streamwise velocity; (c) Reynolds stress component $\overline {u'v'}$; (d) TKE.}
\label{fig:8-profile}
\end{figure}

Figure~\ref{fig:8-profile} shows the profiles of the time-averaged streamwise velocity (a), the root mean square (rms) of the streamwise velocity (b), the $\overline {u'v'}$ component of the Reynolds stress (c) and the TKE (d) at $z/d$ = 2, $y/d$ = 0, $x/d = $ 2, 3, 4, 5 and 6 respectively. As shown in that figure, our findings are consistent with the visualisation results but more intuitive. Furthermore, this figure clearly demonstrates the shrinkage of the recirculation zone and the suppression of flow separation. In Fig.~\ref{fig:8-profile}(a), the mean velocity of the control case proves to be more significant than that of the no-control case, indicating the reduction in the recirculation area. Figure~\ref{fig:8-profile}(b) and (d) exhibits a double-peaked symmetric structure. The induced separated flow behind the model increases the fluctuating velocity at $x/d$ = 2, which is in turn decreased in the downstream region. In case 3, with the control of both plasma actuators, the fluctuating velocity reachs its minimum value in the control cases, indicating it is the optimal control case.\par

\begin{figure}
\centering 
\includegraphics[angle=0, trim=0 0 0 0, width=1\textwidth]{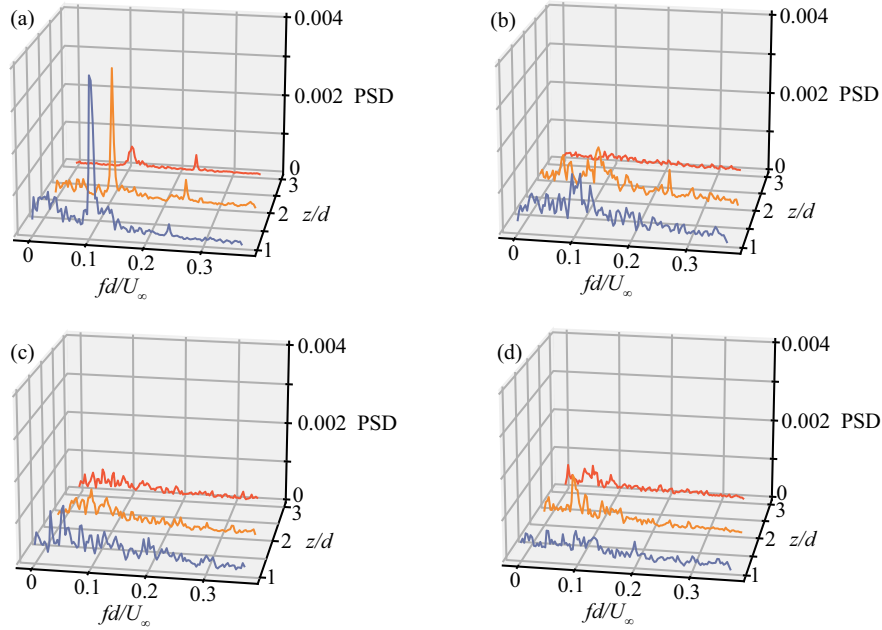}
\caption[]{The power spectral density (PSD) of the streamwise fluctuating velocity $u$ at $x/d$ = 1, $y/d$ = 1, $z/d$ = 1, 2, 3. (a) Uncontrolled case; (b) case 1; (c) case 2; and (d) case 3.}
\label{fig:9-PSD}
\end{figure}

As shown in Fig.~\ref{fig:9-PSD}, the behaviour of the vortex shedding frequency is examined by means of performing a power spectral density (PSD) analysis of the streamwise fluctuating velocity at $x/d$ = 1 and $y/d$ = 1 for different wall-normal locations along the height of the bluff body and different configurations of controlling actuators. The PSD peak indicates the dimensionless vortex shedding frequency, Strouhal number ($St$) = 0.107, which is similar to the ones found in the literature \cite{wang2022}. The $St$ is defined as $fd/U_\infty$, where $f$ is the shedding frequency. In the case of no-control, a secondary peak appears at around $St$ = 0.24 due to the second harmonic.\cite{wang2009} As shown in Fig.~\ref{fig:9-PSD}(a), the strength of the PSD becomes lower with increasing height. At $z/d$ = 3, the PSD decreases rapidly due to the strong vortex shedding strong at the free end. The separated vortices determine the peak of PSD, and the downwash flow from the free end vortex shedding weakens the vortex shedding as a result of suppressing the strength of the PSD. In addition, Fig.~\ref{fig:9-PSD}(b-d) shows the PSD obtained by the three control cases, in which the dominant peak of the PSD exhibites by the no-control case is now decreased, forming certain small peaks of lower strength, indicating the suppression of the periodic vortex shedding behaviour. As shown in Fig.~\ref{fig:9-PSD}(b), the PSD at $z/d$ = 3 has the minimum quantity and best control effect, suggesting that the plasma actuator on the top surface could efficiently strengthen the downwash flow and weaken the spanwise vortex at high elevations.\par

\subsection{Aerodynamic performance of plasma actuators}

\begin{figure}
\centering 
\includegraphics[angle=0, trim=0 0 0 0, width=1\textwidth]{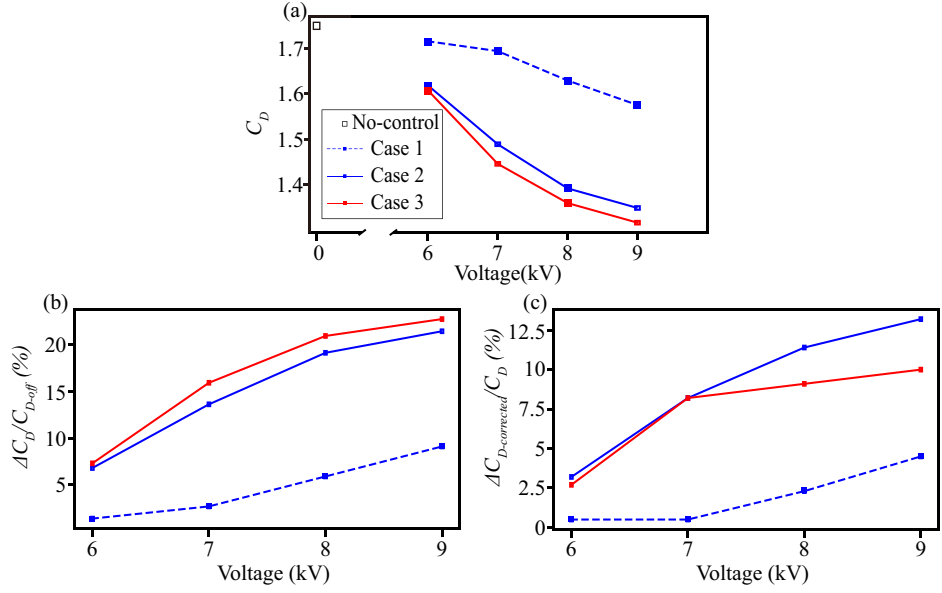}
\caption[]{(a) Drag coefficient; (b) total reduced drag percentages; and (c) corrected reduced drag percentage of the no-control case and three control cases at ${Re}_d$ = 4700.}
\label{fig:10-CD}
\end{figure}

The effect of flow separation can be highlighted in terms of aerodynamic forces acting on the bluff body. Therefore, this section underlines the effect of plasma actuators on the aerodynamic performance (i.e. drag reduction). A load cell is used to perform the drag force measurements at a Reynolds number of approximately 4,700. Here, the near-wake flow structure shows similar behaviour to ${Re}_d$ = 500 but more accessible to enable more proper measurements. As shown in Fig.~\ref{fig:10-CD}(a), the drag coefficient in the uncontrolled case ($C_{D-off}$) is approximately 1.72. After activating the control, the value of $C_D$ is found to decrease significantly in all three cases. In particular, case 3 shows the best performance. It can be observed that the value of $C_D$ in all cases decreases with increasing the input voltage due to the higher induced flow velocity and greater momentum change.\par

Figure~\ref{fig:10-CD}(b) demonstrates the percentage of drag reduction ($\mathrm{\Delta}C_D/C_{D-off}$). In the figure, the maximum drag reduction ratio reaches a value of approximately 9.1$\%$ for case 1, 21.4$\%$ for case 2 and 22.7$\%$ for case 3. Based on these findings, it can be concluded that case 3, which employs both plasma actuators, exhibits a superior performance at high input voltage compared to case 2, where only the rear surface plasma actuator is used. Notably, as the input voltage decreases, the drag reduction in both cases eventually converged, which could be attributed to the insufficient induced flow from the top surface plasma actuator at a low input voltage, which fails to promote the reattachment of the separated flow on the surface.\par

Moreover, we use the corrected drag coefficient to eliminate the thrust effect caused by the bluff body, as shown in Fig. 9(c). The $\Delta{C}_{D-corrected}$ is defined as:

\begin{equation} \label{eqn:eq2}
\Delta{C}_{D-corrected}=\frac{\Delta F_D-T}{\frac{1}{2}\rho U^2_{\infty}A},
\end{equation}

\noindent where $\Delta{F}_D$ is the drag force reduction and $T$ is the thrust from the plasma actuators measured in the quiescent air. The observed drag reduction can be attributed to two different parts: the thrust in a direction opposite from the drag and the reattachment behaviour after flow separation which yielded a decrease in the pressure difference. Furthermore, $\Delta{C}_{D-corrected}$ can correct the drag coefficient after eliminating the thrust effect. As shown in Fig.~\ref{fig:10-CD}(c), at input voltage of 9 $kV$, the $\Delta{C}_{D-corrected}$ for case 1, 2 and 3 are 4.5$\%$, 13.2$\%$ and 10$\%$, respectively. The $\Delta{C}_{D-corrected}$ for case 1 is low compared with the value obtained from the other two cases, suggesting that the thrust is the main contributor of the observe drag reduction. Moreover, $\Delta{C}_{D-corrected}$ in case 3 is lower than that in case 2, which is due to the effect of the top surface plasma actuator.\par

\section{Conclusions}\label{sec:Conclusions}

The present study aims to experimentally investigate the plasma actuator effect on the near-wake flow structure and aerodynamic performance of a wall-mounted square cylinder with an aspect ratio of four at a low Reynolds number. The cylinder is equipped with plasma actuators on the top and rear surfaces, and three control cases are used to enhance the control efficiency of 3-D wake structure. The near-wake structure is studied both qualitatively and quantitatively by means of visualisation, profile of mean velocity, fluctuating velocity, Reynolds shear stress, turbulent kinetic energy and power spectral density analysis.\par

The visualisation results show that all three control cases can suppress the flow separation from the sharp leading edges of the bluff body, with control case 3 (having both plasma actuators) performing exceptionally well. The plasma actuator on the top surface of the bluff body causes the reattachment behaviour observed at the free end, while the plasma actuator on the rear surface vastly reduces the recirculation zone behind the bluff body. The reattachment point moves from $x/d$ = 4.2 in the uncontrolled case to $x/d$ = 2 at the control case 3, which exhibits the best control efficiency.

Furthermore, the turbulent kinetic energy results reveal that the plasma actuators can effectively reduce the value of turbulent kinetic energy downstream, which can reach 67$\%$ after controlling both plasma actuators. However, the chaotic-induced flow of plasma actuators is found to slightly increases turbulent kinetic energy upstream. The statistical results are consistent with the visualisation results, indicating that all control cases have contributed to the suppression of the flow separation.\par

Finally, the power spectral density analysis also showed that all three control cases can weaken the vortex shedding, with the plasma actuator on the top surface strengthening the downwash flow. The drag force measurements demonstrate that all control cases facilitate drag reduction, with the maximum reduction of 9.1$\%$, 21.4$\%$ and 22.7$\%$ in the three controlled cases, respectively. The control case 3 exhibits the optimal drag reduction performance within various driving voltages.\par

Overall, our findings show that case 3, which is controlled by both plasma actuators, is the most effective approach in suppressing flow separation and reducing drag force. This is because the induced flow from the plasma actuators can efficiently exchange momentum, accelerating the flow in the recirculation zone and reducing the  drag force.\par

\begin{acknowledgments}
This work was supported by 'Human Resources Program in Energy Technology' of the Korea Institute of Energy Technology Evaluation and Planning (KETEP), granted financial resource from the Ministry of Trade, Industry \& Energy, Republic of Korea (no. 20214000000140). In addition, this work was supported by the National Research Foundation of Korea (NRF) grant funded by the Korea government (MSIP) (no. 2019R1I1A3A01058576). This work was also supported by the National Supercomputing Center with supercomputing resources including technical support (KSC-2022-CRE-0282).
\end{acknowledgments}

\appendix

\nocite{*}
\bibliography{my-bib}

\end{document}